# Superconductivity in Se-doped new materials $EuSr_2Bi_2S_4F_4$ and $Eu_2SrBi_2S_4F_4$

Zeba Haque[#], Gohil S Thakur[#$], Rainer Pöttgen[§], Ganesan Kalai Selvan[‡], Rangasamy Parthasarathy[#], Sonachalam Arumugam[‡], Laxmi Chand Gupta[#†] and Ashok Kumar Ganguli[#,€,*]

[#]Department of Chemistry, Indian Institute of Technology, New Delhi, India, 110016
[$]Max-Planck-Institute for Chemical Physics of Solids, Dresden, Germany, 01187

[§] Institut für Anorganische und Analytische Chemie, Universität Münster, Corrensstrasse 30, D-48149 Münster, Germany

[‡]Centre for High Pressure Research, School of Physics, Bharathidasan University, Tiruchirapalli, India 620024

[€]Institute of Nano Science & Technology, Habitat Centre, Punjab, India, 160062

**ABSTRACT:** From our powder x-ray diffraction pattern, electrical transport and magnetic studies we report the effect of isovalent Se substitution at S-sites in the newly discovered systems $EuSr_2Bi_2S_4F_4$ and $Eu_2SrBi_2S_4F_4$.We have synthesized two new variants of 3244—type superconductor with Eu replaced by Sr which is reported elsewhere [Z. Haque *et al*]. We observe superconductivity at $T_c$ 2.9 K (resistivity) and 2.3 K (susceptibility) in $EuSr_2Bi_2S_{4-x}Se_xF_4$ series for x = 2. In the other series $Eu_2SrBi_2S_{4-x}Se_xF_4$, two materials (x= 1.5; $T_c$ = 2.6 K and x = 2; $T_c$ = 2.75 K) exhibit superconductivity.

After the discovery of superconductivity in layered $BiS_2$ based compound $Bi_4O_4S_3$ [1], tremendous amount of work on $BiS_2$ based compounds has been carried out in the past four years. Soon after the realization of superconductivity in Bi-O-S system superconductivity was observed in $LaO_{0.5}F_{0.5}BiS_2$ at 2.5 K (ambient pressure) and 10.6 K (high pressure annealing)[2]. Subsequently several groups investigated various other BiS2 based (namely LnOBiS2, AFBiS2, Bi-O-S system) materials [3-4]. All these materials share same basic crystal structure having the conducting $BiS_2$ bilayers alternating with blocking layer of edge sharing $Ln_2O_2$ / $Sr_2F_2$ / $Eu_2F_2$ / $Bi_2O_2$ tetrahedra. Along with the substitution at rare earth / Eu,Sr- / Bi- / O-site in the blocking layer [5-6] there have been successful attempts to modify the conducting $BiS_2$ layers by introducing isovalent selenium atoms [7-11]. Superconductivity in EuBiS2F [12] (1121 – type superconductor) and $Eu_3Bi_2S_4F_4$ [13] (3244 - type superconductor) at 0.3 and 1.5 K, was reported by H. F. Zhai *et al* without any external doping. $Eu_3Bi_2S_4F_4$ has a layered tetragonal structure (SG: *I*4/*mmm*, *Z* = 2), which consist of two (Eu,Sr)FBiS₂ subcell interspersed with an extra $EuF_2$ layer [13]. It is remarkable to note here that in our attempt to synthesize $EuBiS_{2-x}Se_xF$ (x = 1) we ended up with multi phase binary and ternary products however $Eu_3Bi_2S_{4-x}Se_xF_4$ ($0 \le x \le 2$) forms [10]. With the motivation to induce superconductivity in the two new materials synthesized and characterized by us earlier $EuSr_2Bi_2S_4F_4$ and $Eu_2SrBi_2S_4F_4$ [14], we have substituted Se at S-sites and carried out their resistivity and magnetic properties in the substituted samples at ambient pressure.

Polycrystalline samples with nominal composition $EuSr_2Bi_2S_{4-x}Se_xF_4$ (x = 0, 0.5, 1, 1.5, 2) and $Eu_2SrBi_2S_{4-x}Se_xF_4$ (x = 0, 0.5, 1, 1.5, 2) were synthesized by solid state reaction method. The reactants $EuF_3$, $EuF_2$, $SrF_2$, $Bi_2S_3$, $Bi_2Se_3$ and Bi metal powder were thoroughly mixed, pelletized and heated in evacuated quartz ampoule at 850°C for 35 hours. Thus obtained dark black pellets remained stable in air for several weeks. The precursors $Bi_2S_3$ and $Bi_2Se_3$ used in the reaction were presynthesized by heating stoichiometric amounts of Bi metal powder with S and Se powder respectively at 500°C for 12 hours. Phase purity of all the samples was checked by powder x-ray diffraction with Cu-*Kα* radiation using a *Bruker* D8 advance diffractometer. The temperature dependence of resistivity and magnetization measurements (2 – 300 K) were performed on a *Cryogen free* Physical Properties Measurements System- Vibrating Sample Magnetometer (PPMS-VSM).

The powder X-ray diffraction data shown in figure 1 (a and b) confirms the formation of nearly single phase tetragonal $Eu_3Bi_2S_4F_4$ crystal structure with space group *I*4/*mmm*. Small amounts of $Bi_2S_3$, $Bi_2Se_3$ and $EuF_2$ impurity phases were also detected. As expected, the substitution of larger Se atom at S-site causes a systematic expansion of lattice volume. Inset of figure 1 (a and b) shows lattice expansion upon Se doping in both the systems.

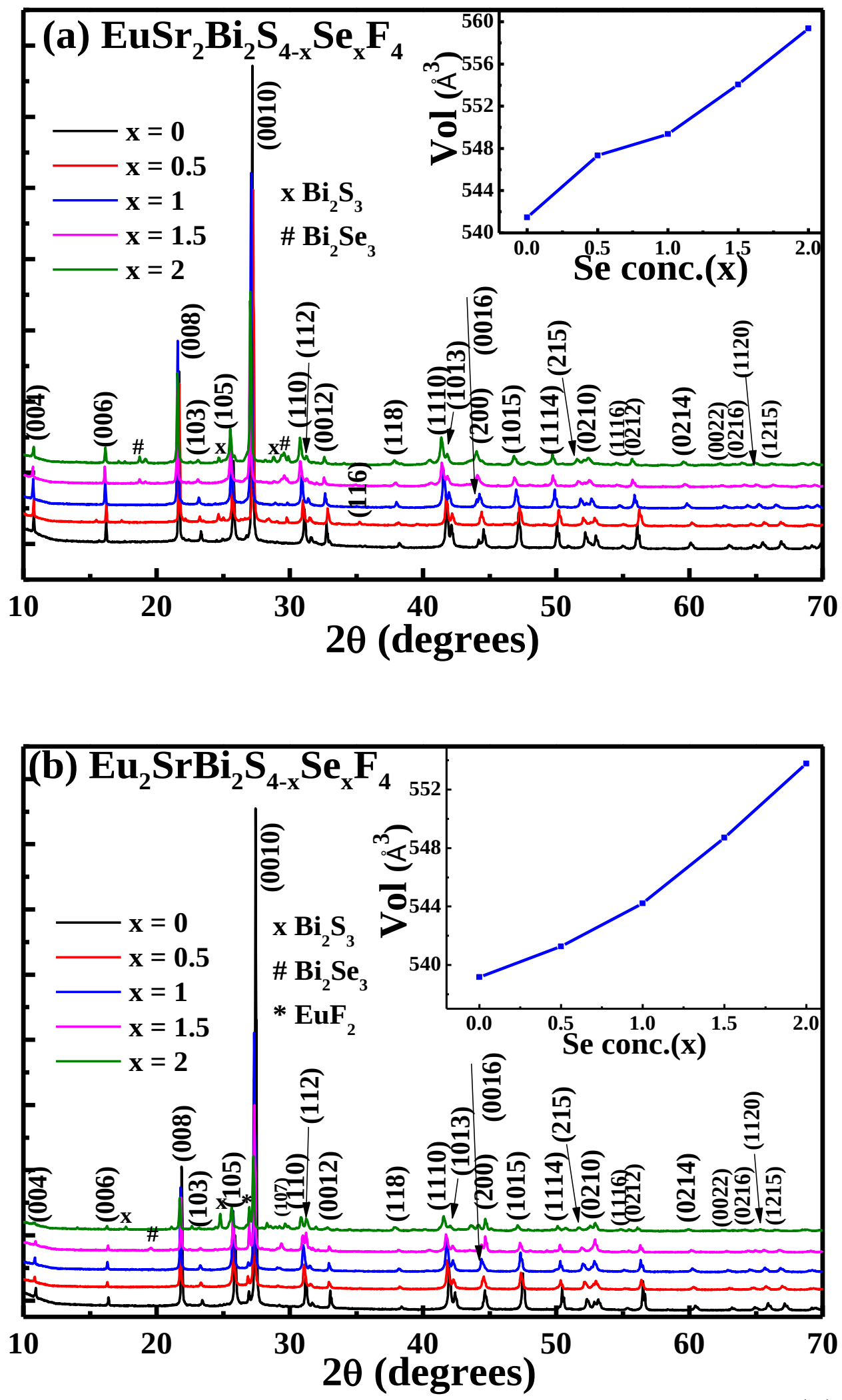

Figure 1. Powder X-ray diffraction pattern of (a) $EuSr_2Bi_2S_{4-x}Se_xF_4$ (x = 0, 0.5, 1, 1.5, 2) and (b) $Eu_2SrBi_2S_{4-x}Se_xF_4$ (x = 0, 0.5, 1, 1.5, 2) showing the formation of tetragonal 3244 phase. Insets show expansion of lattice volume in both the systems.

The temperature dependent resistivity studies on $EuSr_2Bi_2S_{4-x}Se_xF_4$ (x = 0, 1, 1.5, 2) are shown in Fig.2(a). The Se free compound in the $EuSr_2Bi_2S_4F_4$ series shows a semimetallic behavior with an upturn in resistivity below 100 K as shown in figure 2a and the value of resistivity is quite close to that of the parent $Eu_3Bi_2S_4F_4$ [13]. We see a decrease in normal state resistivity upon Se substitution. For x = 0, we observe a broad hump in resistivity shown by arrows in figure 2a, at 250 K. This feature shifts towards a lower temperature upon Se-doping and vanishes for x = 2.0. This compound $EuSr_2Bi_2S_2Se_2F_4$ (x = 2.0) shows metallic conduction and a sharp superconducting transition with $T_c^{onset}$ = 2.9 K (inset of figure 2a).

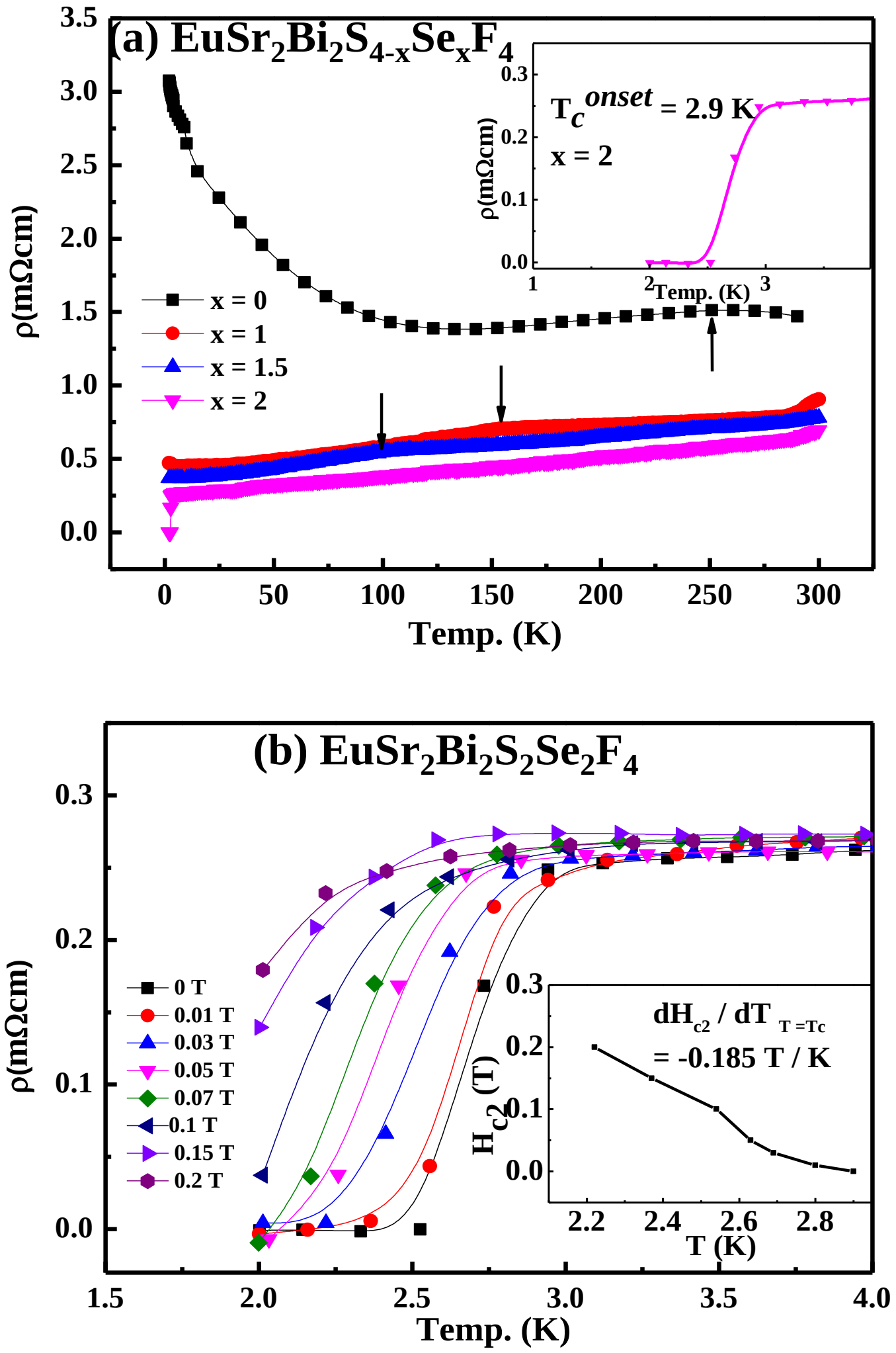

Figure 2. (a) Variable temperature resistivity ρ (T) of $EuSr_2Bi_2S_{4-x}Se_xF_4$ (x = 0, 1, 1.5, 2). The inset shows sharp superconducting drop at 2.9 K in $EuSr_2Bi_2S_2Se_2F_4$. (b) Field dependent resistivity ρ (T) for $EuSr_2Bi_2S_2Se_2F_4$ in the superconducting region. Inset shows $H_{c2}$ vs T plot.

The transition width ΔT ~ 0.4 K indicates good quality of the superconducting phase. The field dependence of resistive $T_c$ for the superconducting sample, $EuSr_2Bi_2S_2Se_2F_4$, is shown in figure 2b. On applying a field, $T_c$ shifts to lower temperature. The upper critical field of $EuSr_2Bi_2S_2Se_2F_4$ obtained in these studies is shown in the inset of figure 2b. *Hc*2(0) is calculated by using WHH formula and is estimated to be 3.72 kOe. This value of *Hc*2(0) is much lower than that in $Eu_3Bi_2S_4F_4$ [13]. Figure 3a shows the variable temperature resistivity on $Eu_2SrBi_2S_{4-x}Se_xF_4$ (x = 0, 1.5, 2). The Se free compound, in $Eu_2SrBi_2S_4F_4$ series, shows a semimetallic behavior, similar to $EuSr_2Bi_2S_4F_4$ with an upturn in resistivity value below 100 K. Among the Se doped samples we observe that the normal state resistivity value decreases upon Se substitution. We observe nearly temperature independent resistivity above 20 K in $Eu_2SrBi_2S_{2.5}Se_{1.5}F_4$ and a metallic conduction for $Eu_2SrBi_2S_2Se_2F_4$. Both these compounds exhibit superconducting transition at 2.6 and 2.7 K respectively (shown in figure 3b). The field dependent resistivity as a function of temperature for $Eu_2SrBi_2S_{2.5}Se_{1.5}F_4$ is shown in figure 3c. Superconducting transition dies off on applying field. We estimated $H_{c2}(0)$ 4.74 kOe in $Eu_2SrBi_2S_{2.5}Se_{1.5}F_4$.

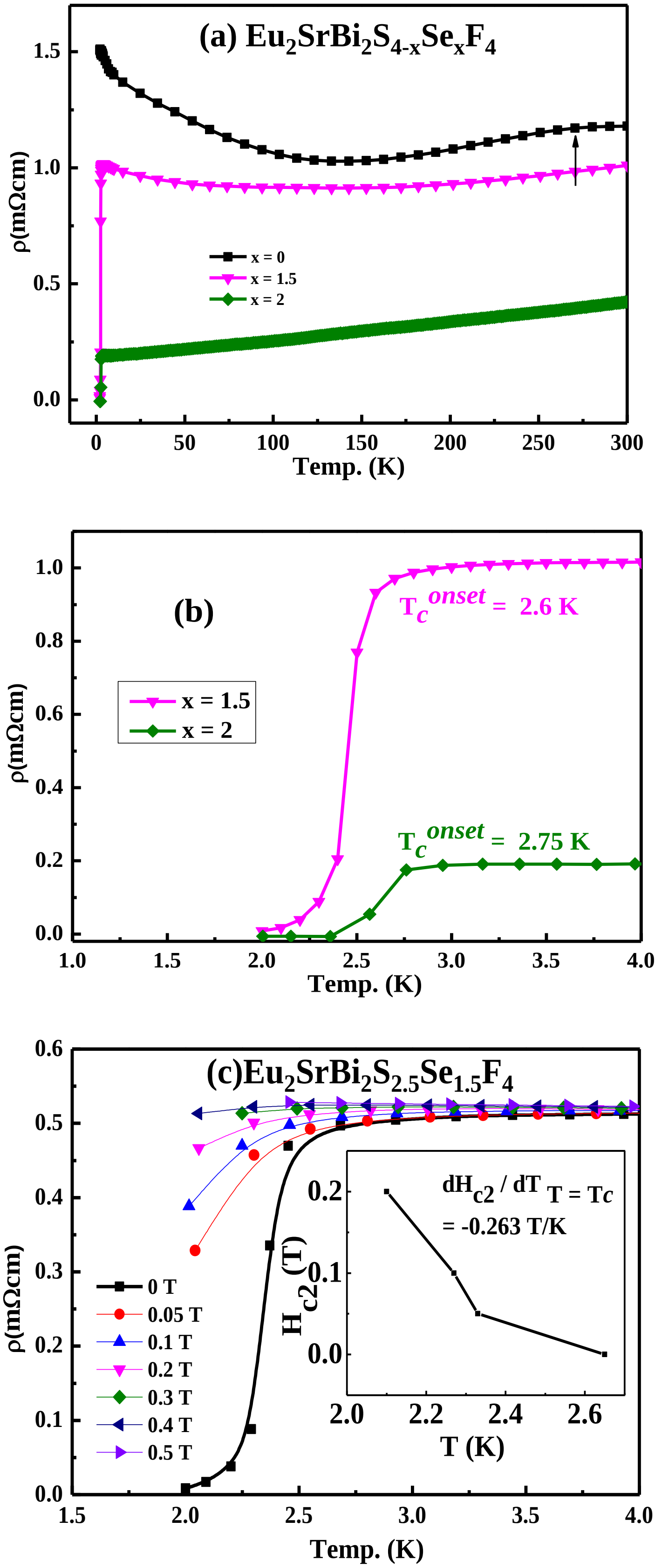

Figure 3. (a) Variable temperature resistivity ρ (T) for $Eu_2SrBi_2S_{4-x}Se_xF_4$ (x = 0, 1.5, 2). (b) Enlargement of onset of the superconducting transition at low temperatures for x= 1.5 and 2. (c) Field dependent resistivity ρ (T) for $Eu_2SrBi_2S_{2.5}Se_{1.5}F_4$ in the superconducting region. Inset shows $H_{c2}$ vs T plot.

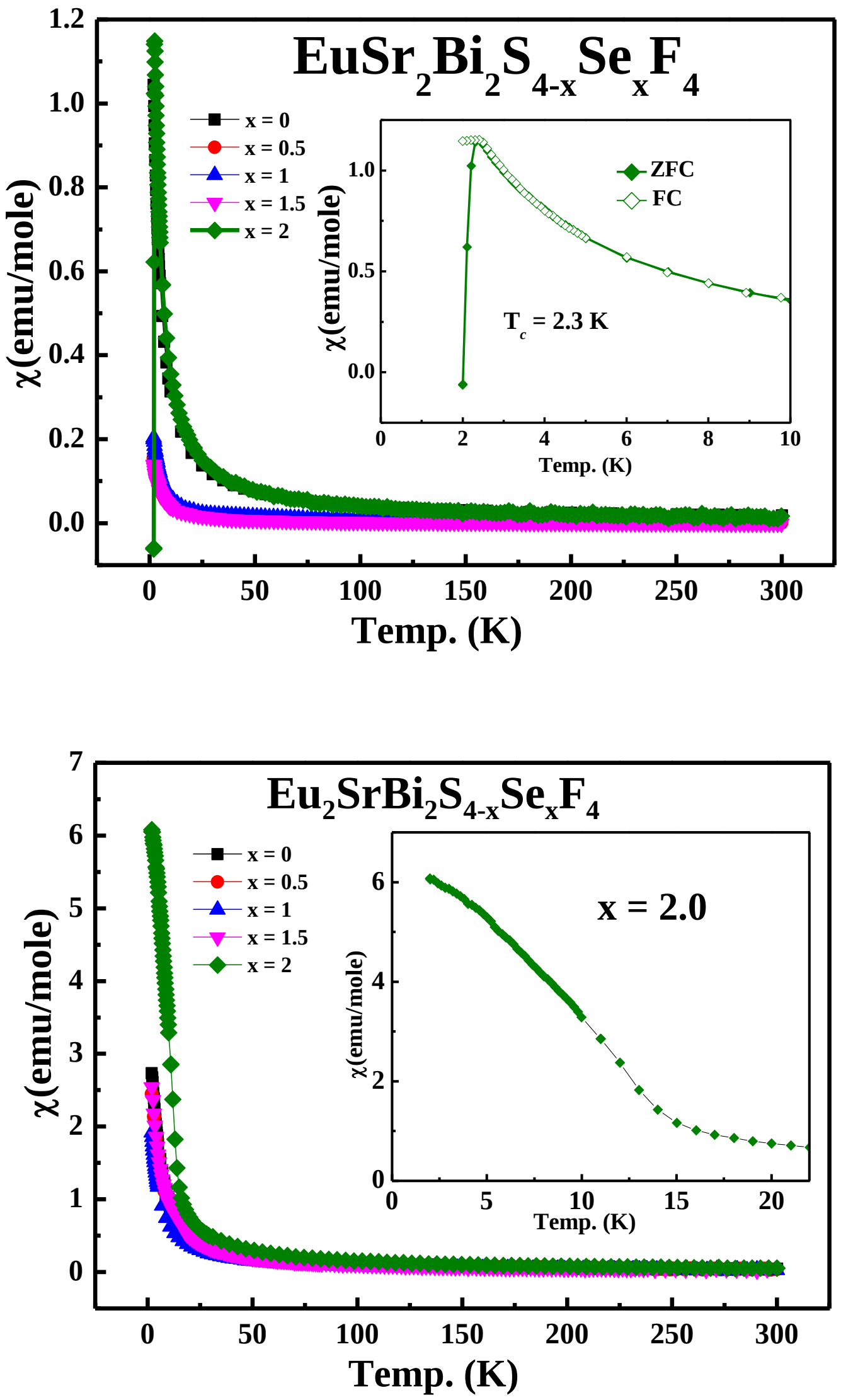

Figure 4. DC susceptibility measurement as a function of temperature (χ)T for (a) $EuSr_2Bi_2S_{4-x}Se_xF_4$ (x = 0, 0.5, 1, 1.5, 2). Inset shows sharp drop in χ at 2.3 K in ZFC for $EuSr_2Bi_2S_2Se_2F_4$ which indicates onset of SC and (b) $Eu_2SrBi_2S_{4-x}Se_xF_4$ (x = 0, 0.5, 1, 1.5, 2). Inset shows ferromagnetic ordering ~14 K for $Eu_2SrBi_2S_2Se_2F_4$.

DC magnetization studies as a function of temperature in an applied field of 20 Oe for both the series $EuSr_2Bi_2S_{4-x}Se_xF_4$ (x = 0, 0.5, 1, 1.5, 2) and $Eu_2SrBi_2S_{4-x}Se_xF_4$ (x = 0, 0.5, 1, 1.5, 2) is shown in figure 4(a and b). In $EuSr_2Bi_2S_{4-x}Se_xF_4$ series, sample with x = 2.0 sample shows a drop in $\chi$ at 2.3 K. This indicates the onset of superconductivity as observed in the resistivity data. The other samples of this series show paramagnetic behavior down to 2 K. In the second series $Eu_2SrBi_2S_{4-x}Se_xF_4$ (x = 0, 0.5, 1, 1.5, 2) we see a paramagnetic behavior of all the samples down to 2 K. Diamagnetic signal is not seen for x = 1.5 and 2 samples, which may be due to the presence of some magnetic interactions in the compounds. In the undoped compound $Eu_2SrBi_2S_4F_4$, one of the Eu ions is essentially divalent as shown in our Mossbauer result [14]. The addition of Se is not expected to change the valence state of Eu because this substitution is isovalent. The paramagnetic contribution of divalent Eu ion overrides the diamagnetic signal which accounts for the absence of diamagnetic signal in $Eu_2SrBi_2S_{2.5}Se_{1.5}F_4$ and $Eu_2SrBi_2S_2Se_2F_4$.

In conclusion we have synthesized new superconductors by Se-substitution at S-sites in $EuSr_2Bi_2S_4F_4$ and $Eu_2SrBi_2S_4F_4$. We report superconductivity in $EuSr_2Bi_2S_2Se_2F_4$; $T_c$ 2.9 K (resistivity) and 2.3 K (susceptibility). In the other series $Eu_2SrBi_2S_{4-x}Se_xF_4$, two materials (x= 1.5 and 2) exhibit superconductivity at $T_c$ = 2.6 and 2.7 K respectively. Absence of diamagnetic signal in $Eu_2SrBi_2S_{2.5}Se_{1.5}F_4$ and $Eu_2SrBi_2S_2Se_2F_4$ is explained by the diamagnetic contribution of the divalent Eu ions.

AUTHOR INFORMATION

Corresponding Author

Present Address

A.K.G.: Director, Institute for Nano Science and Technology, Phase 10, sector 64, Mohali, Punjab 160062, India.

†L.C.G.: Visiting scientist at Solid State and Nanomaterials Research Laboratory, Department of Chemistry, IIT Delhi, India.

ACKNOWLEDGMENT

A.K.G. acknowledges DST for financial support. Z.H. and G.S.T. acknowledge UGC and CSIR, respectively, for fellowships. G.K.S. and S.A. acknowledge the DST(FISTPURSE, & SERB), DRDO, CEFIPRA, BRNS and UGC (RFSMS- SRF) for their financial support.